\documentstyle[12pt,epsf]{article}

\def\bz{{\bf z}}

\def\bs{{\bf s}}

\def\bL{{\bf L}}

\newcommand{\be}{\begin{equation}}
\newcommand{\ee}{\end{equation}}
\newcommand{\bea}{\begin{eqnarray}}
\newcommand{\eea}{\end{eqnarray}}

\newcommand{\lan}{\langle}
\newcommand{\ran}{\rangle}

\newcommand{\p}{\partial}
\begin{document}

\vskip 2cm
\begin{center}
\Large
{\bf  Wilson loop evaluations } \\
{\bf in the stochastic vacuum model } \\
\vskip 0.5cm
\large
 Ken Williams \\
{\small \em Department of Physical and Earth Sciences,  } \\
{\small \em Jacksonville State University, Jacksonville, Alabama 36265 } \\
\end{center}
\thispagestyle{empty}
\vskip 0.7cm

\begin{abstract}
The stochastic vacuum model description of a heavy meson is
discussed in the context of a gauge-invariant approach where Wilson
loop expectation values appear naturally in the O($v^2$) spin-orbit Hamiltonian. These expectation values have been derived elsewhere, however by a procedure whose legitimacy is now placed in question.  Here they are derived by standard functional methods with a result that is identical to the previous one. In addition, a full spin-independent Hamiltonian reduction to O($v^2$) is carried out.
\end{abstract}

\newpage

\section{introduction}

It was the work of Leutwyler and Voloshin some years ago in the
context of the sum rule formalism that first suggested the fundamental
nonlocality of nonperturbative interactions between hadronic quarks \cite{lv}. Leading effects were later shown proportional to a
gluon condensate \cite{md} thereby excluding the possibility of a
purely local description.  In a separate line of development Wilson's
lattice work \cite{wilson} led to the well-known area law as a
qualitative formulation of color confinement. There, gauge invariance
of the hadronic state is the guiding principle. There too what is
fundamental is the presence of a nonlocal operator - the Wilson loop.

In the stochastic vacuum model (svm) of Dosch and Simonov \cite{ds} we
find these salient features mutually complementary, where area
law asymptotics follow from a non-zero gluon condensate.  Active gluon
degrees via the field's correlation length measured against $ Q\bar{Q}
$ correlations also come into the picture; this in such a way that
the model also accounts for intermediate as well as short-range
perturbative behavior.

The aim of the present article is to describe the reduction of the svm
to an effective $ O(v^2) $ spin-orbit interaction Hamiltonian for the heavy $
Q\bar{Q} $ state. In fact, a general reduction in terms of Wilson loop
expectation values (ev) has already been made \cite{bram}, thus
leaving as the main focus of this study only the ev derivations
themselves.  The ev too have been derived \cite{sim,kus,bv,kn,sd}, however by a procedure whose legitimacy is here placed in question: From the functional variation of the Wilson-loop
\bea
\delta \imath \ln W &=&   \int \delta\sigma_{\mu\nu}(z_1) \lan\lan
 F_{\mu\nu}(z_1) \ran\ran \label{three1} 
\eea
is presumed the functional derivative relation
\bea
\frac{\delta}{\delta\sigma_{\mu\nu}(z_1)} i\ln W &=& \lan\lan F_{\mu\nu}(z_1) \ran\ran  \label{four1}
\eea
resulting in a spin-orbit interaction that e.g. satisfies the Gromes relation. 

We argue that procedure (1) $\to $ (2) lacks legitimacy by virtue of insufficient degrees of freedom characterizing the variational area element $\delta\sigma_{\mu\nu} $. In section 2 the familiar svm spin-orbit Hamiltonian is derived along the
usual lines from assumption (1) $\to$ (2). We then demonstrate the general invalidity of the assumption and show that when omitted it leaves the Hamiltonian
insufficiently specified. Additional constraints are derived in
section 3 where the model statement itself is expanded to
$ O(m^{-2}) $ and velocity coefficients matched term by
term. This procedure leads again to the familiar svm spin-orbit Hamiltonian; I.e, our spin-orbit result agrees with the one derived from (2), although, as we point out, generally applied (2) may lead to incorrect results \cite{ken1}. Finally, we lay out details of a full reduction to the $ O(m^{-2}) $ spin-independent Hamiltonian.

\section{  functional variation and the svm  }

From a Foldy-Wouthysen reduction of the gauge invariant $Q\bar{Q}
$ 4-point function the $O( m^{-2} )$ interaction Hamiltonian of
\cite{bram} is given in terms of the Wilson loop
\bea
V &=& V_{so} + V_{si} \nonumber \\
V_{so} &=& \sum_{i=1}^2 (-1)^{i} \frac{1}{2m_i} \epsilon_{j \mu\nu k}
s_i^j \dot{z}_i^\mu
\lan\lan F^{\nu k}(z_i) \ran\ran \nonumber \\
&\equiv & \left( \frac{1}{2m^2_1} \bL_1 \cdot \bs_1 - \frac{1}{2m^2_2} \bL_2 \cdot \bs_2 \right) [V_0(r) +2V_1(r)]^\prime/r \nonumber \\
&& + \quad
\frac{1}{m_1m_2} (\bL_1 \cdot \bs_2 - \bL_2 \cdot \bs_1 )
 V_2^\prime (r)/r \label{one} \\
\int dt V_{si} &=& i \ln W = i\ln \frac{1}{3} \lan tr P
\exp( ig\oint dt ( A_0 - \dot{z}^i A^i ) ) \ran \nonumber
\eea
where Darwin and hyperfine terms, not relevant to the present
discussion, are omitted. Key to this approach therefore is the
evaluation of the six independent expectation values $ \lan\lan
F_{\mu\nu} \ran\ran $. The defining Euclidean svm statement in the
present context is \cite{sim}
\bea
\ln W &=& - \frac{\beta}{2} \int d\sigma_{\mu\nu}(u) \int
 d\sigma_{\lambda\rho}(v) 
\bigg\{(\delta_{\mu\lambda}\delta_{\nu\rho} - \delta_{\mu\rho}
\delta_{\nu\lambda} ) D(w^2) \label{two} \\ && \frac{1}{2} \left[
\frac{\p}{\p u_\mu}(w_\lambda \delta_{\nu\rho} - w_\rho \delta_{\nu\lambda})
+\frac{\p}{\p u_\nu}(w_\rho \delta_{\mu\lambda} - w_\lambda \delta_{\mu\rho})
\right] D_1(w^2) \bigg\} , \, w\equiv u-v \nonumber
\eea
where the integrals are evaluated over instantaneous straight-line
surfaces: $ ( u_i = s z_{1i} + (1-s) z_{2i} , u_4 = t ), \, 0 \leq s
\leq 1 $, with $ d\sigma_{\mu\nu}(u) \equiv dt ds ( \p u_\mu /dt)(\p
u_\nu /ds) $. $ \beta $ is the gluon condensate, and D and $ D_1 $ are
gluon correlation functions that fall off rapidly. From the quark
world-line variation, $ \delta z_{1\mu} $, of the Wilson loop
\bea
\delta \imath \ln W &=&   \int \delta\sigma_{\mu\nu}(z_1) \lan\lan
 F_{\mu\nu}(z_1) \ran\ran \label{three} \\
&=& -i \beta \int \delta\sigma_{\mu\nu}(z_1) \int
 d\sigma_{\lambda\rho}(v)
\bigg\{(\delta_{\mu\lambda}\delta_{\nu\rho} - \delta_{\mu\rho}
\delta_{\nu\lambda} ) D(\tilde{w}^2) \nonumber \\ && \frac{1}{2} \left[
\frac{\p}{\p z_{1\mu}}(\tilde{w}_\lambda \delta_{\nu\rho} - \tilde{w}_\rho \delta_{\nu\lambda})
+\frac{\p}{\p z_{1\nu}}(\tilde{w}_\rho \delta_{\mu\lambda} - \tilde{w}_\lambda \delta_{\mu\rho})
\right] D_1(\tilde{w}^2) \bigg\} , \, \tilde{w} \equiv z_1-v \nonumber
\eea
is proposed in references \cite{sim,kus,bv} the following the
functional relation
\bea
\frac{\delta}{\delta\sigma_{\mu\nu}(z_1)} i\ln W &=& \lan\lan F_{\mu\nu}(z_1) \ran\ran  \label{four}
\eea
where $ \delta \sigma_{\mu\nu} \equiv (dz_\mu \delta z_\nu - dz_\nu
\delta z_\mu )/2 $ is called the variational area element.
From this follows
\bea
\lan\lan F_{0l}(z_1) \ran\ran &=&  \beta r_l \int d\tau \left\{\int_0^r
d\lambda \frac{1}{r} D(\tau^2+\lambda^2)+\frac{1}{2} D_1(\tau^2+r^2) \right\}
\label{five} \\
\lan\lan F_{il}(z_1) \ran\ran &=&  \beta (\dot{z}_{1l}r_i -\dot{z}_{1i}r_l )
\int d\tau \int_0^r d\lambda \frac{1}{r}\left(1-\frac{\lambda}{r} \right)
D(\tau^2+\lambda^2) \nonumber \\ &&
+ \beta (\dot{z}_{2l}r_i -\dot{z}_{2i}r_l )
\int d\tau \left\{ \int_0^r d\lambda \frac{\lambda}{r^2} D(\tau^2+\lambda^2)
+\frac{1}{2} D_1(\tau^2 + r^2) \right\} \nonumber
\eea
yielding spin-orbit potentials
\bea
V_0^\prime(r) &=& \beta \int d\tau \left[ \int^r_0 d\lambda
D(\tau^2 +\lambda^2 )+\frac{r}{2} D_1(\tau^2 +r^2 ) \right] \nonumber \\
V_1^\prime(r) &=& - \beta \int d\tau \int^r_0 d\lambda \left(
1-\frac{\lambda}{r} \right) D(\tau^2 +\lambda^2 ) \label{six} \\
V_2^\prime(r) &=& \beta \int d\tau \left[ \int^r_0 d\lambda \frac{\lambda}{r} D(\tau^2 +\lambda^2 )+\frac{r}{2} D_1(\tau^2 +r^2 ) \right] \nonumber
\eea
which together e.g. satisfy the relation of Gromes \cite{gromes}
\bea
 [V_0 + V_1 - V_2]^\prime &=& 0  . \label{seven}
\eea
We show here that variational derivative (\ref{four}) does indeed not follow from variation (\ref{three}). One should notice that in passing from (\ref{three}) to (\ref{four}), (1) to (2), is the assumption that
there are in the variational area element at least six quark and
six antiquark coordinate degrees of freedom - an assumption clearly in error. 

The Wilson loop is a functional of quark and antiquark world lines. With
this in mind we rewrite the rhs of (\ref{three})
\bea
\int dt \delta z_{1\nu} \dot{z}_{1\mu} \lan\lan F_{\mu\nu}(z_1)
 \ran\ran &=& \int dt \delta z_{1\nu} \dot{z}_{1\mu} \Theta_{\mu\nu} \,
\label{eight}
\eea
where
\bea
\Theta_{\mu\nu} &\equiv & -i \beta  \int
 d\sigma_{\lambda\rho}(v)
\bigg\{(\delta_{\mu\lambda}\delta_{\nu\rho} - \delta_{\mu\rho}
\delta_{\nu\lambda} ) D(\tilde{w}^2) \label{nine} \nonumber \\ && \frac{1}{2} \left[
\frac{\p}{\p u_\mu}(\tilde{w}_\lambda \delta_{\nu\rho} - \tilde{w}_\rho \delta_{\nu\lambda})
+\frac{\p}{\p u_\nu}(\tilde{w}_\rho \delta_{\mu\lambda} - \tilde{w}_\lambda \delta_{\mu\rho})
\right] D_1(\tilde{w}^2) \bigg\} \, . \nonumber
\eea
From this we should like to extract the six element $ \{ \lan\lan F^{\mu\nu} \ran\ran \} $ for determination of the Hamiltonian (\ref{one}). The set appears in (\ref{one}) as the linearly transformed 
\bea
\{ \delta_{ij}
\delta_{\mu\nu} \dot{z}_{1\mu} \lan\lan F_{\nu j}(z_1) \ran\ran ,
\epsilon_{ij\mu\nu} \dot{z}_{1\mu} \lan\lan F_{\nu j}(z_1) \ran\ran \}
\label{set}
\eea
- a set of vectors spanning the space of field tensor elements. The
first subset accounts for spin independent $ V_{si} $, by Stokes
theorem, and the second for spin dependent $ V_{so} $ directly. It
should be clear that the first subset appears in
(\ref{three}). Hence independence of $ V_{so} $ with respect to variation
(\ref{three}) follows from the linear independence of vectors in the
complete set (\ref{set}). The independence is immediately
established from the nonsingularity of the set's
coefficient matrix
\[ \det \left(
\begin{array}{cccccc}
1 & 0 & 0 & -\dot{z}_2 & -\dot{z}_3 & 0 \\
0 & 1 & 0 & \dot{z}_1 & 0 & -\dot{z}_3 \\
0 & 0 & 1 & 0 & \dot{z}_1 & \dot{z}_2 \\ 
-\dot{z}_2 & \dot{z}_1 & 0 & 2 & 0 & 0 \\
-\dot{z}_3 & 0 & \dot{z}_1 & 0 & 2 & 0 \\ 
0 & -\dot{z}_3 & \dot{z}_2 & 0 & 0 & 2
\end{array}
\right) \quad \sim \quad (2-\dot{z}^2)^2 \]
where the vectors have been arranged by row with field tensor elements
in ascending order, left to right.

For concreteness, then, variation (\ref{three})
\bea
\int dt \delta z_{1 4} \dot{z}_{1j} \lan\lan F_{4j}(z_1)
 \ran\ran &=& \int dt \delta z_{1 4} \dot{z}_{1j} \Theta_{4j} \label{11} \\
\int dt \delta z_{1j} (\lan\lan F_{4j}(z_1) \ran\ran +  \dot{z}_{1i}
 \lan\lan F_{ij}(z_1)
 \ran\ran) &=& \int dt \delta z_{1j} ( \Theta_{4j} + \dot{z}_{1i} 
\Theta_{ij} ) \nonumber
\eea
leads to
\bea
\lan\lan F_{0j}(z_1) \ran\ran &=& \Theta_{0j} + (\dot{z}_1^2 \hat{r}_j
-\dot{z}_{1j} \dot{z}_{1i} \hat{r}_i ) f_1^\prime(r) + 
 (\dot{z}_{1i} \dot{z}_{2i} \hat{r}_j
-\dot{z}_{2j} \dot{z}_{1i} \hat{r}_i ) f_2^\prime(r) \nonumber \\
\lan\lan F_{ij}(z_1) \ran\ran &=& \Theta_{ij} +
(\hat{r}_i \dot{z}_{1j} -\hat{r}_j \dot{z}_{1i}) f_1^\prime(r) +
(\hat{r}_i \dot{z}_{2j} -\hat{r}_j \dot{z}_{2i}) f_2^\prime(r) \label{12}
\eea
yielding when installed into (\ref{one}) the spin-orbit potentials
\bea
V_1^\prime(r) &=&  f_1^\prime(r) -
 \beta \int d\tau \int^r_0 d\lambda \left(
1-\frac{\lambda}{r} \right) D(\tau^2 +\lambda^2 ) \label{13} \\
V_2^\prime(r) &=& - f_2^\prime(r)
+ \beta \int d\tau \left[ \int^r_0 d\lambda \frac{\lambda\
}{r} D(\tau^2 +\lambda^2 )+\frac{r}{2} D_1(\tau^2 +r^2 ) \right] \nonumber
\eea
for arbitrary functions $ (f_1^\prime, f_2^\prime ) $. I.e., solutions (\ref{12}) and (\ref{13}) satisfy variation (\ref{three}), yet leaves the spin-orbit Hamiltonian $ V_{so} $ entirely unspecified. This
claim may be verified by direct substitution of (\ref{12}) into (\ref{three})and (\ref{12}) into (\ref{one}).

\section{specification of the svm Wilson loop expectation values }

We now revisit the spin-orbit Hamiltonian derivation beginning again from (\ref{three}).    
As in ref\cite{ken1}, additional constraints are found from the functional
expansion of the model's defining statement in orders of heavy quark
velocity. Accordingly, definition (\ref{two}) is expanded
\bea
\imath \ln W  &\simeq & \int dt dt^\prime \bigg\{ V^0(r,r^\prime) +
 (\dot{z}_{1i }+\dot{z}_{2i}) V^\alpha_i(r,r^\prime) \label{141} \\ && + 
 (\dot{z}_{1i}  \dot{z}_{1j}^\prime + \dot{z}_{2i}  \dot{z}_{2j}^\prime ) V^\beta_{ij}(r,r^\prime) +
(\dot{z}_{1i } \dot{z}_{2j}^\prime + \dot{z}_{2i}  \dot{z}_{1j}^\prime) V^\gamma_{ij}(r,r^\prime)
\bigg\} . \nonumber
\eea
which on the O($v^2$) approximation 
\bea
r_i^\prime &\approx& r_i - \dot{r}_i \tau + \ddot{r}_i \tau^2/2 \, , \qquad \tau = t - t^\prime \label{eval}
\eea
yields
\bea
\imath \ln W  &\simeq & \int dt \bigg\{ V_0(r) +
 (\dot{\bz}_1^2 +\dot{\bz}_2^2) V_a(r) + 
 \dot{\bz}_1 \cdot \dot{\bz}_2 V_b(r) \label{14} \\ && +
[ (\dot{\bz}_1 \cdot \hat{r})^2 +(\dot{\bz}_2 \cdot \hat{r})^2 ] V_c(r) +
( \dot{\bz}_1 \cdot \hat{r} ) ( \dot{\bz}_2 \cdot \hat{r} ) V_d(r)
\bigg\} . \nonumber
\eea
for functions $ V^\alpha, V^\beta, V^\gamma, V_0, V_a, V_b, V_c, V_d $  to be determined.
Equation (\ref{141}) leads to ( see appendix )
\bea
\lan\lan F_{ij}(z_1) \ran\ran &=&  \beta (\dot{z}_{1j}r_i -\dot{z}_{1i}r_j )
\int d\tau \int_0^r d\lambda \frac{1}{r}\left(1-\frac{\lambda}{r} \right)
D(\tau^2+\lambda^2) \label{15} \\ &&
+ \beta (\dot{z}_{2j}r_i -\dot{z}_{2i}r_j )
\int d\tau \left\{ \int_0^r d\lambda \frac{\lambda}{r^2} D(\tau^2+\lambda^2)
+\frac{1}{2} D_1(\tau^2 + r^2) \right\} \, ,\nonumber
\eea
and for the spin-independent potentials of (\ref{14}) we find upon application of (\ref{eval}) (see appendix)
\bea
V_0(r) &=& \beta  \int d\tau \int_0^r d\lambda \bigg[
\left( r-\lambda  \right) D(\tau^2 +\lambda^2 ) +
\frac{\lambda}{2}  D_1 (\tau^2 +\lambda^2 ) \, \bigg]  \label{16} \\
V_a(r) &=&  - \beta  \int d\tau \int_0^r d\lambda \bigg[
\left( \frac{r}{6} -\frac{\lambda}{4} +\frac{\lambda^3}{12 r^2}
+\frac{\lambda \tau^2 }{2 r^2} -\frac{\tau^2}{2 r}
   \right) D(\tau^2 +\lambda^2 )\nonumber \\ && \qquad \qquad
+\left( \frac{\lambda}{8} -\frac{\lambda^2}{4 r} +\frac{\lambda^3}{8 r^2}
-\frac{\lambda \tau^2 }{4 r^2} +\frac{\tau^2}{4 r}
     \right)  D_1 (\tau^2 +\lambda^2 ) \, \bigg]  \\
V_b(r) &=& - \beta  \int d\tau \int_0^r d\lambda \bigg[
\left( \frac{r}{6} -\frac{\lambda^3}{6 r^2}
-\frac{\lambda \tau^2 }{ r^2} +\frac{\tau^2}{ r}
    \right)  D(\tau^2 +\lambda^2 ) \nonumber \\ && \qquad \qquad \quad
+\left( \frac{\lambda^2}{2 r} -\frac{\lambda^3}{4 r^2}
+\frac{\lambda \tau^2 }{2 r^2} -\frac{\tau^2}{2 r}
     \right) D_1(\tau^2 +\lambda^2 ) \, \bigg]  \\
V_c(r) &=& - \beta  \int d\tau \int_0^r d\lambda \bigg[
\left( -\frac{r}{6} +\frac{\lambda^2}{2 r}
-\frac{\lambda^3 }{3 r^2}    \right)  D(\tau^2 +\lambda^2 ) \bigg] \\
V_d(r) &=& - \beta  \int d\tau \bigg[ \int_0^r d\lambda
\left(-\frac{r}{6} -\frac{\lambda^2}{ r}
+\frac{2 \lambda^3 }{3 r^2}    \right)  D(\tau^2 +\lambda^2 )
-\frac{r^2}{4} D_1(\tau^2 + r^2) \bigg] \label{20} 
\eea
A couple observations: The ev  of (\ref{15}) follow from strict independence of operators $ (r,v)$ in the order expansion (\ref{141}). The independence is relaxed in the O($v^2$) evaluation (\ref{eval}) that leads to spin-independent potentials (\ref{16}) - (\ref{20}). This explains the discrepancy between  ev (\ref{15}) and the one obtained in an earlier effort by the present author \cite{unrevised} where instead the ev are derived improperly from expansion (\ref{14}). We note that as (\ref{15}) is identical with (\ref{five}) it yields upon insertion into (\ref{one}) the spin-orbit potentials (\ref{six}) that satisfy relation (\ref{seven}). However, only incidentally \cite{ken}. We also note that in the long-range regime the spin-independent interaction, (\ref{16}) - (\ref{20}), agrees with minimum area or flux-tube asymtotics \cite{bram, ken1}
\bea
V_{si} &\to & - \frac{1}{6} a \frac{L^2}{r} \left( \frac{1}{m_1^2}+\frac{1}{m_1^2}-\frac{1}{m_1 m_2} \right) \, , \\   && \qquad \qquad \qquad \qquad a \equiv \beta\int d\tau\int_0^r d\lambda D(\tau^2 +\lambda^2)  \, . \nonumber
\eea
This too is a correction to the aforementioned earlier effort where in the corresponding equation (\ref{14}) the relation
\bea
\dot{\bz}^2_{Euclidean} &=& - \dot{\bz}^2_{Minkowskian}
\eea
is overlooked. In ref.\cite{bv} a spin independent Hamiltonian
reduction of the svm to $ O(v^2) $ is carried out. The apparent disagreement with potentials (\ref{16}) - (\ref{20}) is due to the use of a partial integration scheme slightly different from the one applied here. The two results are equivalent. 

\section{appendix}

From the functional identities
\bea
\frac{\delta \xi(t) }{\delta \xi(t^\prime) } &=& \delta(t-t^\prime) \\
\delta f(\xi) &=& \delta \xi \, ( \frac{\p}{\p \xi } f(\xi) )
\eea
the Taylor function and functional expansions are carried out to
$ O(v^2) $
\bea
f(\dot{\bz}_1,\dot{\bz}_2) &=& f(0)+\dot{z}_1^i \int dt^\prime \left(
\frac{\delta }{\delta \dot{z}_1^i } f(\dot{\bz}_1^\prime,\dot{\bz}_2) 
\right)_{0} + \dot{z}_2^i \int dt^\prime \left(
\frac{\delta }{\delta \dot{z}_2^i } f(\dot{\bz}_1,\dot{\bz}_2^\prime )
\right)_{0} \nonumber \\ && +
\frac{1}{2} \dot{z}_1^i \dot{z}_1^j \int \int dt^\prime dt^{\prime
\prime } \left( \frac{\delta }{ \delta \dot{z}_1^i } \frac{\delta }{
\delta \dot{z}_1^{\prime j }} f(\dot{\bz}_1^{\prime \prime },\dot{\bz}_2)
\right)_{0} \nonumber \\ && +
\frac{1}{2} \dot{z}_2^i \dot{z}_2^j \int \int dt^\prime dt^{\prime
\prime } \left( \frac{\delta }{ \delta \dot{z}_2^i } \frac{\delta }{
\delta \dot{z}_2^{\prime j }} f(\dot{\bz}_1,\dot{\bz}_2^{\prime \prime })
\right)_{0}\nonumber \\ && +
\dot{z}_1^i \dot{z}_2^j \int \int dt^\prime dt^{\prime
\prime } \left( \frac{\delta }{ \delta \dot{z}_1^i } \frac{\delta }{
\delta \dot{z}_2^j } f(\dot{\bz}_1^\prime ,\dot{\bz}_2^{\prime \prime })
\right)_{0}  \, + h.o. \\
F[\dot{\bz}_1,\dot{\bz}_2] &=& F[0]+\int dt \dot{z}_1^i \left(
\frac{\delta }{\delta \dot{z}_1^i } F[\dot{\bz}_1^\prime,\dot{\bz}_2]
\right)_{0} + \int dt \dot{z}_2^i \left(
\frac{\delta }{\delta \dot{z}_2^i } F[\dot{\bz}_1,\dot{\bz}_2^\prime ]
\right)_{0} \nonumber \\ && +
\frac{1}{2} \int \int dt dt^\prime \dot{z}_1^i \dot{z}_1^{j\prime}
 \left( \frac{\delta }{ \delta \dot{z}_1^i } \frac{\delta }{
\delta \dot{z}_1^{\prime j }} F[\dot{\bz}_1^{\prime \prime },\dot{\bz}_2]
\right)_{0} \nonumber \\ && +
\frac{1}{2} \int \int dt dt^\prime \dot{z}_2^i \dot{z}_2^{j\prime}
 \left( \frac{\delta }{ \delta \dot{z}_2^i } \frac{\delta }{
\delta \dot{z}_2^{\prime j }} F[\dot{\bz}_1,\dot{\bz}_2^{\prime \prime }]
\right)_{0} \nonumber \\ && +
\int \int dt dt^\prime \dot{z}_1^i \dot{z}_2^j
 \left( \frac{\delta }{ \delta \dot{z}_1^i } \frac{\delta }{
\delta \dot{z}_2^j } F[\dot{\bz}_1^\prime ,\dot{\bz}_2^{\prime \prime }]
\right)_{0}  \, + h.o.
\eea
where primes indicate time dependence, e.g., $ \xi^\prime = \xi(t^\prime) $ ,
and subscript `` 0'' means evaluation at $ \dot{\bz}_1 =
\dot{\bz}_2 = 0 $. The spatial field tensor Wilson loop expectation
value to first order is then
\bea
\lan \lan F_{ij}(z_1) \ran \ran &=& \lan \lan F_{ij} \ran \ran_{0} +
\dot{z}_{1k} \int dt^\prime \left( \frac{\delta }{ \delta  \dot{z}_{1k} } 
\lan \lan F_{ij} \ran \ran^\prime \right)_0 +
\dot{z}_{2k} \int dt^\prime \left( \frac{\delta }{ \delta  \dot{z}_{2k} }
\lan \lan F_{ij} \ran \ran^\prime \right)_0 \nonumber \\
&=& \lan \lan \p_i A_j \ran \ran_0 - \lan \lan \p_j A_i \ran \ran_0 \nonumber \\&&+ \dot{z}_{1k} \int dt^\prime \left( \frac{\delta }{ \delta \dot{z}_{1k} } 
\lan \lan \p_{i} A_j \ran \ran^\prime_1 -\frac{\delta }{\delta  \dot{z}_{1k} } 
\lan \lan \p_j A_i \ran \ran^\prime_1 \right)_0 \nonumber \\ && +
\dot{z}_{2k} \int dt^\prime \left( \frac{\delta }{ \delta  \dot{z}_{2k} }
\lan \lan \p_i A_j \ran \ran^\prime_1 - \frac{\delta }{ \delta  \dot{z}_{2k} }
\lan \lan \p_j A_i \ran \ran^\prime_1 \right)_0 \, . \label{qq}
\eea
To find the above rhs we expand both sides of the svm, (\ref{two}), to O($v^2$)
\bea
\imath \ln W &\simeq & \imath \ln W_0  - \int dt \left( \dot{z}_{1i} \lan \lan A_i(z_1) \ran \ran_0 + \dot{z}_{2i} \lan \lan A_i(z_2) \ran \ran_0 \right) \nonumber \\&&
- \frac{1}{2} \int dt dt^\prime \left[
 \dot{z}_{2i} \dot{z}_{1j}^\prime
 \left( \frac{\delta }{ \delta  \dot{z}_{2j} }
\lan \lan A_j(z_1) \ran \ran^\prime \right)_0 - \dot{z}_{1i} \dot{z}_{2j}^\prime \left(
 \frac{\delta }{ \delta  \dot{z}_{1i} }
\lan \lan A_j(z_2) \ran \ran^\prime \right)_0 \right] \nonumber \\ &&
-\frac{1}{2} \int dt dt^\prime \left[ \dot{z}_{1i} \dot{z}_{1j}^\prime
\left( \frac{\delta }{ \delta  \dot{z}_{1i} }
\lan \lan A_j(z_1) \ran \ran^\prime \right)_0 - \dot{z}_{2i} \dot{z}_{2j}^\prime
\left( \frac{\delta }{ \delta  \dot{z}_{2i} }
\lan \lan A_j(z_2) \ran \ran^\prime \right)_0 \right] \nonumber \\
 &=&\int dt dt^\prime \bigg\{ V^0(r,r^\prime) +
 (\dot{z}_{1i} +\dot{z}_{2i}) V^\alpha_i(r,r^\prime) \nonumber \\ && + 
 (\dot{z}_{1i}  \dot{z}_{1j}^\prime + \dot{z}_{2i } \dot{z}_{2j}^\prime) V^\beta_{ij}(r,r^\prime) +
(\dot{z}_{1i } \dot{z}_{2j}^\prime + \dot{z}_{2i } \dot{z}_{1j}^\prime) V^\gamma_{ij}(r,r^\prime)
\bigg\} \label{qqq}
\eea
where we find
\bea
V^\alpha_i &=& -i\beta \int_0^1 ds ds^\prime s (\omega_i r_j r_j^\prime -\omega_j r_j r_i^\prime ) \tau \frac{\p}{\p \tau^2} D_1 \\
V^\beta_{ij} &=&  -i \frac{\beta}{2} \int_0^1 ds ds^\prime s s^\prime \bigg\{ (\delta_{ij}r_k r_k^\prime - r_i^\prime r_j ) ( D + D_1) \\ && \times
[ \omega_k \omega_l r_k r_l^\prime \delta_{ij}+\omega_i\omega_j r_k r_k^\prime -\omega_i\omega_k r_j r_k^\prime - \omega_j\omega_k r_k r_i^\prime] \frac{\p}{\p \tau^2} D_1 \bigg\} \nonumber \\
V^\gamma_{ij} &=&  -i \frac{\beta}{2} \int_0^1 ds ds^\prime (1-s) s^\prime \bigg\{ (\delta_{ij}r_k r_k^\prime - r_i^\prime r_j ) ( D + D_1) \\ && \times
[ \omega_k \omega_l r_k r_l^\prime \delta_{ij}+\omega_i\omega_j r_k r_k^\prime -\omega_i\omega_k r_j r_k^\prime - \omega_j\omega_k r_k r_i^\prime] \frac{\p}{\p \tau^2} D_1 \bigg\} \nonumber
\eea
and where
\bea
\frac{\delta }{ \delta  \dot{z}_{1j} } \imath \ln  W &=&
\frac{\delta }{ \delta  \dot{z}_{1j} } \imath \ln \frac{1}{3}
 \lan  tr P \exp[ \imath g
\int dt^\prime ( A_0 +\dot{z}_i A_i ) ]  \ran \nonumber \\
&=&  - \lan \lan A_j(z_1) \ran \ran
\eea
has been used. From this follows (equating velocity coefficients)
\bea
\lan \lan  A_i(z_1) \ran \ran_0 &=& - \int dt^\prime V^\alpha_i (r,r^\prime) \nonumber \\
\int dt^\prime \left(\frac{\delta }{ \delta  \dot{z}_{1i }}\lan \lan A_j(z_1) \ran \ran^\prime \right)_0 &=& - 2 \int dt^\prime V^\beta_{ij}(r,r^\prime) \nonumber \\
\int dt^\prime \left(\frac{\delta }{ \delta  \dot{z}_{2i }}\lan \lan A_j(z_1) \ran \ran^\prime \right)_0 &=& - 2 \int dt^\prime V^\gamma_{ij}(r,r^\prime) \, .
\eea
This result installed into (\ref{qq}) yields the ev (\ref{15}) upon performing the indicated differentiations.

For the spin-independent Hamiltonian we make the further O($v^2$) reduction on the rhs of (\ref{qqq})
\bea
r_i^\prime &\simeq& r_i - \dot{r}_i \tau + \ddot{r}_i \tau^2/2 \nonumber \\
D(\omega^2)&\simeq& \left( 1+O_{ii}\frac{\p}{\p \tau^2}+O_{ii}^2 \frac{\p}{\p \tau^2}\frac{\p}{\p \tau^2} /2 \right) D(\tau^2+\lambda^2 r^2) 
 \eea
where
\bea
O_{ij} &\equiv& \lambda\tau(r_i\dot{v}_j+r_j\dot{v}_i) +
\tau^2\dot{v}_i\dot{v}_j -\lambda\tau^2(r_i\ddot{v}_j+r_j\ddot{v}_i)/2 \nonumber \\
u_i &=& s z_{1i} +(1-s) z_{2i} \nonumber \\
v_i &=& s^\prime z_{1i} +(1-s^\prime) z_{2i}\nonumber \\
\lambda &\equiv& s-s^\prime \, , \, \tau \equiv t - t^\prime 
\eea
which leads from (\ref{qqq}) to the spin-independent potentials (\ref{16}) - (\ref{20}) upon performing the necessary summations and partial integrations.


\begin{thebibliography}{}


\bibitem{lv}
H. Leutwyler, Phys.\ Lett.\ {\bf B98}, 447 (1981) ;
M.B. Voloshin, Nucl.\ Phys.\ {\bf B187}, 365 (1981).

\bibitem{md}
U. Marquard and H.G. Dosch, Phys.\ Rev.\ {\bf D35}, 2238 (1987).

\bibitem{wilson}
K.G. Wilson, Phys.\ Rev.\ {\bf D10}, 2445 (1974).

\bibitem{ds}
H.G. Dosch and Yu. A. Simonov, Phys.\ Lett.\ {\bf B205}, 339 (1988).

\bibitem{bram}
N.~Brambilla,
P. Consoli, and G.~M.~Prosperi, Phys.\ Rev. \ {\bf D50}, 5878 (1994).

\bibitem{sim}
Yu. A. Simonov, Nucl.\ Phys.\ {\bf B324}, 67 (1989).

\bibitem{kus}
V. M. Kustov, Phys.\ Atom.\ Nucl.\ {\bf 60}, 1754 (1997).

\bibitem{bv}
N. Brambilla and A. Vario, Phys.\ Rev.\ {\bf D55}, 3974 (1997).

\bibitem{kn}
Yu. S. Kalashnikova and A. V. Nefediev, Phys.\ Lett.\ {\bf B414}, 149 (1997).

\bibitem{sd}
M. Schiestl and H.G. Dosch, Phys.\ Lett.\ {\bf B209}, 85 (1988).

\bibitem{ken1}
K. Williams, {\em The minimum area, the flux tube, and Thomas precession }
 (hep-ph/ 9806269).


\bibitem{gromes}
D. Gromes, Z.\ Phys.\ {\bf C26}, 401, (1984).

\bibitem{ken}
K. Williams, {\em Revisiting the Eichten-Feinberg-Gromes $ Q\bar{Q} $
  Spin-Orbit Interaction } (hep-ph/ 9607211).


\bibitem{ovw}
M.G.Olsson, S.Veseli and K.Williams, Phys.\ Rev.\ {\bf D53 }, 4006 (1996).

\bibitem{bp}
A. Barchielli, E. Montaldi and G. M. Prosperi, Nucl.\ Phys.\ {\bf B296},
625 (1988).

\bibitem{unrevised}
See earlier version of present article.


\end{thebibliography}
\end{document}